\documentstyle[12pt]{article}
\documentstyle{a4wide}
\newcommand{\vgl}[1]{eq.(\ref{#1})}
\newcommand{\scri}{{\cal S}}
\newcommand{\exscri}{{\cal S}_\eta}
\newcommand{\nice}{C^\infty(\scri)}
\newcommand{\exnice}{C^\infty(\exscri)}
\newcommand{\der}{Der(\nice)}
\newcommand{\exder}{Der(\exnice)}

\newsavebox{\uuunit}
\sbox{\uuunit}
    {\setlength{\unitlength}{0.825em}
     \begin{picture}(0.6,0.7)
        \thinlines
        \put(0,0){\line(1,0){0.5}}
        \put(0.15,0){\line(0,1){0.7}}
        \put(0.35,0){\line(0,1){0.8}}
       \multiput(0.3,0.8)(-0.04,-0.02){12}{\rule{0.5pt}{0.5pt}}
     \end {picture}}

\newcommand{\half}{{\textstyle\frac{1}{2}}}
\newcommand{\lder}{\stackrel{\leftharpoonup}{\partial }\!}
\newcommand{\rder}{\stackrel{\rightharpoonup}{\partial }\!}
\newcommand{\etab}{\bar\eta}

\begin{document}
\begin{flushleft}
\LaTeX
\end{flushleft}
\begin{flushright}
KUL-TF-93/38\\
hep-th/9309080\\
September 1993.
\end{flushright}
\begin{center}
{\large\bf A noncommutative note on the}\\
{\large\bf antibracket formalism.}\\
\vskip 1cm
\small
\begin{center}
{\bf F. Vanderseypen}$^\sharp$\\
\vskip 1cm
Instituut voor Theoretische Fysica
        \\Katholieke Universiteit Leuven
        \\Celestijnenlaan 200D
        \\B-3001 Leuven, Belgium\\[0.3cm]
        E-mail : Swa\%tf\%fys@cc3.kuleuven.ac.be\\
        $\sharp$ Aspirant NFWO, Belgium\\
\end{center}
\vskip 1cm
{\bf Abstract }
\end{center}
\vskip .5cm
\footnotesize
 {\sl We introduce a noncommutative calculus on the
odd-symplectic superspace $\scri$ of fields and antifields. To this end
we have to extend $\scri$ to $\exscri$ by including an extra anticommuting
field $\eta$. As a consequence we show that
the commutator induced on $\exnice\times T^\star(\exnice)$ is
proportional
to the antibracket. The $\Delta$-operator is an element of the quotient
space of derivations twisted by the antibracket $A\der$ and $\der$. The
natural measure on $\exscri$
is shown to be invariant under canonical transformations provided a certain
'wave equation' is satisfied.}\\
\normalsize
\vskip 1cm
\noindent
\underline{{\sf Introduction.}}\\
Recently, it was shown that noncommutative geometry is a unifying framework
for describing various theories. Continuum, lattice and certain
q-deformed theories were recovered by applying the noncommutative
calculus \cite{qdeform,lattice} and with the same ease one can elaborate
the formalism to include
the Ito or Stratonovich calculus \cite{beast}. The basic (underlying)
structure in
these developments is a commutative algebra of functions (equivalent to a
finite-dimensional manifold) whereon a noncommutative set of rules is
superimposed.\\
It is a remarkable fact that the very same method applied on a superspace
(which is nothing but a special case of noncommutative (background) space),
supplied with an odd-symplectic structure, results in the antibracket
formalism \cite{bvleuv,Batman,BV,bvsb,Schw,Witt}. So, loosely speaking,
we can say that
the antibracket formalism is a noncommutative calculus on superspace.\\
In the following we introduce an extra field $\eta$ on the
(infinite-dimensional) space of fields and antifields. This extra field is
the analog of the 'evolution parameter' or extra 'time' in \cite{beauty}.
We refer to this article for a full discussion. Although in our context we
cannot give an acceptable meaning to this $\eta$-parameter we will derive
a certain 'wave equation' in the extended superspace and this seems to
confirm the 'time'-analogy.
\vskip 1cm
\noindent
\underline{{\sf The antibracket derived.}}\\
Let $\scri$ be the superspace of fields and antifields
\begin{equation}
\scri=\{\phi^\alpha =\{\phi^A,\phi_A^\star\}\mid A=1,\ldots,N\, ;
\:\epsilon (\phi_A^\star)=\epsilon (\phi^A )+1\},
\end{equation}
where $\epsilon $ is the Grassmann parity
\begin{equation}
\epsilon (\phi^\alpha )\equiv \epsilon ^\alpha \;\;\;\;\alpha =1,\ldots,2N.
\end{equation}
When no confusion is possible we abreviate $\epsilon ^\alpha $ to $\alpha$.
We denote by
$C^\infty(\scri)$ the ring of nice functions on $\scri$ and the graded
commutator is defined as
\begin{equation}
[F,G\} := F\,G-(-)^{FG}G\, F.
\label{commut}
\end{equation}
Next, we enlarge $\scri$ to
$\exscri:\,=\scri\cup\{\eta\}$ where $\eta$ is an anticommuting field
($\epsilon (\eta)=1$).
We have the following identities :
\begin{equation}
[\phi^\alpha ,\phi^\beta \}=[\phi^\alpha ,\eta\}=0.
\end{equation}
It is easily checked that the following commutation rules on
$\exnice\times T^\star(\exnice)$ form a consistent noncommutative
calculus (cfr. \cite{beauty}) :
\begin{equation}
[\phi^\alpha ,d\phi^\beta\}=\imath\hbar\,d\eta\,\delta^{\alpha \beta },
\label{nckernel}
\end{equation}
\begin{equation}
[\phi^\alpha
,d\eta\}=[\eta,d\eta\}=[\eta,d\phi^\alpha \}=0.
\end{equation}
\noindent
The supersymplectic metric $\delta $ is not
necessarily in the canonical form, this means that $\delta $ can depend on
$\phi$ and $\eta$ \cite{bvleuv,Hata}. Note that $\hbar\longrightarrow 0$
gives the classical de Rham calculus.
Next, I would like to show that \vgl{nckernel} implies the
occurence of the antibracket
in its most general form, i.e. on an arbitrary odd-symplectic space.
To this end we define for $F\in\exnice$
\begin{equation}
dF=d\phi^\alpha \,\rder_\alpha F+d\eta\,\rder_\eta F.
\label{differential}
\end{equation}
Here $\rder_\eta$ and $\rder_\alpha $ are operators acting from the left
on $F$, i.e. \vgl{differential} is the defining relation for  $\rder_\eta$
and $\rder_\alpha $. Actually, we will see below that $\rder_\eta$ is NOT
equal to $\rder /\partial \eta$.
Letting $d$ act on $[F,\phi^\beta \}=0$ yields
\begin{equation}
[F,d\phi^\beta \}=(-)^{F(\beta +1)+\beta }\,[\phi^\beta ,dF\}.
\end{equation}
It is then straightforward to check that :
\begin{equation}
[F,dG\}
= \imath\,\hbar\,d\eta\:(F,G)_{BV},
\label{favorite}
\end{equation}
for functions $F,G\in\exnice$. Earlier, relation (\ref{favorite}) was
shown to
hold \cite{Aristo} in the special case of a finite dimensional superspace.
Using this relation one proves the various identities of the antibracket
\cite{bvleuv}.\\
\vskip .3cm
\noindent
\underline{{\sf The $\Delta$-operator
with density function.}}\\
It is easily seen that $\rder_\alpha \in\exder$ while $\rder_\eta$ is not,
actually
\begin{eqnarray}
d(F\,.G)&=&dF\,.G+(-)^FF.\,dG\nonumber\\
&=&dF\,.G+(-)^{FG}dG\,.F+(-)^F\,[F,dG\},\nonumber\\
\end{eqnarray}
hence
\begin{equation}
\rder_\eta(F\,.G)=\rder_\eta F\,.G +(-)^F F\,.\rder_\eta G +\imath\,
\hbar\,(F,G)_{BV}.
\label{etaLeibnitz}
\end{equation}
And it is then straightforward to check that
\begin{equation}
\rder_\eta :\,=\delta_{der}+\half\imath\, \hbar\Delta
\label{Itoder}
\end{equation}
satisfies \vgl{etaLeibnitz}, where the $\Delta$-operator was introduced
\begin{equation}
\Delta:\,=(-)^A\rder_A\rder^A
\end{equation}
and
$\delta_{der}\in \exder$ is an arbitrary derivation.\\
Operators satisfying \vgl{etaLeibnitz} are called 'twisted derivations' in
the literature \cite{Coq}. So we conclude that the $\Delta$-operator is a
representative
in the space $A\exder$ of derivations twisted by the antibracket,
$A\exder\supset\Delta\bmod\exder$. It is interesting to note here that
\vgl{Itoder} is the analog of the Ito-derivative in the stochastic
calculus \cite{beast}.
One can go a step further, since
$\delta_{der}$ is an arbitrary derivation we can add to the $\Delta$-operator
the following
\begin{equation}
\rder_\eta =\tilde\delta_{der}+\half\imath\,
\hbar\left(\Delta+\half\,(\log\rho,\cdot)\right)
\end{equation}
for $\rho\in\exnice$, $\epsilon (\rho)=0$. Such that we can set
\begin{equation}
\rder_\eta =\tilde\delta_{der}+\half\imath\, \hbar\Delta_\rho,
\end{equation}
where the $\Delta_\rho$ is the $\Delta$-operator appearing in
\cite{Batman,Hata} and $\tilde\delta_{der}\in\exder$.\\
\vskip .3cm
\noindent
\underline{{\sf The wave equation in $\exnice$.}}\\
Let us introduce an antifield $\bar\eta$ for the $\eta$-parameter and set
the extended antibracket equal to
\begin{equation}
(\cdot,\cdot)_{EBV} :\,=(\cdot,\cdot)_{BV}+
\frac{\lder}{\partial \eta}\!\cdot\,
\frac{\rder}{\partial {\etab}}\!\cdot-
\frac{\lder}{\partial {\etab}}\!\cdot\,\frac{\rder}{\partial \eta}\cdot,
\end{equation}
in accordance with the general philosophy of the formalism and
\vgl{commut}.
Next we compute the Berezin determinant for an infintesimal canonical
transformation including $\eta,\etab$ :
\begin{eqnarray}
\phi'^\alpha &=& \phi^\alpha +(\phi^\alpha ,F)_{EBV},\nonumber\\
\eta'&=& \eta + (\eta,F)_{EBV}, \nonumber\\
\end{eqnarray}
and analog transformations for the antifields. Remenber that the
function $F$ has to be odd here. The determinant is
\begin{equation}
Ber\left\{\frac{(\phi',\phi'^\star,\eta',\etab')}{(\phi,\phi^\star,\eta,
\etab )}\right\}= 1-2\Delta F+2\frac{\rder}{\partial
\eta}\!\frac{\rder}{\partial {\etab}}\! F.
\end{equation}
So, it seems natural to extend the phase space to include the pair
$(\eta,\etab)$ and to demand that
\begin{equation}
\Delta F=\frac{\rder}{\partial \eta}\!\frac{\rder}{\partial {\etab}}\! F
\label{waveq}
\end{equation}
which is seen to be the 'wave equation' in extended phase space, if one is
willing to keep the analog between $\eta$ and 'time' as explained before.
Of course, \vgl{waveq} is only a necessary condition for invariance of the
(extended) measure.
\vskip .3cm\noindent
\underline{{\sf Conclusion, queries.}}\\
This short note aimed to show the close connection between the antibracket
formalism and noncommutative geometry. Apart from the epistemological
question of what the meaning of the $\eta$-parameter is, there are many
conceptual problems. It is natural to ask how one could derive the quantum
master-equation in this context, notice that \vgl{waveq} 'looks' very much
like the quantum master-equation. In any case, I am convinced that this
noncommutative approach to the antibracket theory is a promising direction.
\vskip 1cm
\begin{center}
{\sf I wish to thank V. M\"{u}ller-Hoissen and A. Dimakis \\
for encouragement and discussions. \\
I am very grateful for their help.}
\end{center}

\end{document}